\documentclass{elsart}

\usepackage{graphicx}
\usepackage{ifthen}
\usepackage{xspace}
\usepackage{amssymb}
\usepackage{amsmath}
\usepackage{cite}                 
\usepackage{url}

\newcommand{\cer}{\v{C}erenkov\xspace}

\newcommand{\mc}{Monte Carlo\xspace}

\newcommand{\dcs}{doubly Cabibbo suppressed\xspace}

\newcommand{\cf}{Cabibbo favored\xspace}

\newcommand{\GP}{Genetic Programming\xspace}

\newcommand{\lsig}{\ensuremath{\ell/\sigma_\ell}\xspace}

\newcommand{\mathsl}[1]{\mbox{\textsl{#1}}}

\newcommand{\meson}[1]{\ensuremath{\mathsl{#1}}}
\newcommand{\baryon}[1]{\ensuremath{\mathsl{#1}}}

\newcommand{\quark}[1]{\ensuremath{\mathsl{#1}}}
\newcommand{\antiquark}[1]{\ensuremath{\overline{\mathsl{#1}}}}

\newcommand{\boson}[1]{\ensuremath{\mathsl{#1}}}

\newcommand{\dq}{\quark{d}\xspace}

\newcommand{\cq}{\quark{c}\xspace}
\newcommand{\cbq}{\antiquark{c}\xspace}

\newcommand{\proton}{\baryon{p}\xspace}

\newcommand{\pion}{\ensuremath{\pi}\xspace}
\newcommand{\piplus}{\ensuremath{\pion^+}\xspace}
\newcommand{\piminus}{\ensuremath{\pion^-}\xspace}

\newcommand{\kaon}{\ensuremath{\meson{K}}\xspace}

\newcommand{\kminus}{\ensuremath{\kaon^-}\xspace}
\newcommand{\kplus}{\ensuremath{\kaon^+}\xspace}

\newcommand{\kstar}{\ensuremath{\kaon^*(892)^0}\xspace}

\newcommand{\kfttr}{\ensuremath{\kaon^*_0(1430)^0}\xspace}

\newcommand{\dmeson}{\ensuremath{\meson{D}}\xspace}
\newcommand{\dplus}{\ensuremath{\dmeson^{+}}\xspace}

\newcommand{\dzero}{\ensuremath{\dmeson{}^{0}}\xspace}

\newcommand{\dsplus}{\ensuremath{\dmeson_{s}^{+}}\xspace}

\newcommand{\lc}{\ensuremath{\Lambda_c^+}\xspace}

\newcommand{\xicp}{\ensuremath{\Xi_c^+}\xspace}

\newcommand{\wboson}{\ensuremath{\boson{W}}\xspace}

\newcommand{\wplus}{\ensuremath{\wboson^{+}}\xspace}

\newcommand{\kpipi}{\ensuremath{\dplus \to  \kaon^- \piplus \piplus}\xspace}
\newcommand{\kpipidcsd}{\ensuremath{\dplus \to  \kaon^+ \piplus \piminus}\xspace}

\newcommand{\kkpi}{\ensuremath{\dplus \to  \kminus \kplus \piplus}\xspace}

\newcommand{\dskkpi}{\ensuremath{\dsplus \to  \kminus \kplus \piplus}\xspace}
\newcommand{\dskkpidcsd}{\ensuremath{\dsplus \to  \kplus \kplus \piminus}\xspace}

\newcommand{\pkpi}{\ensuremath{\lc \to \proton \kaon^- \piplus}\xspace}
\newcommand{\pkpidcsd}{\ensuremath{\lc \to \proton \kaon^+ \piminus}\xspace}

\newcommand{\gev}{\ensuremath{\mathrm{GeV}}\xspace}

\newcommand{\eqnref}[1]{Eq.~(\ref{eqn:#1})}
\newcommand{\figref}[1]{Figure~\ref{fig:#1}}

\newcommand{\tabref}[1]{Table~\ref{tab:#1}}

\newcommand{\refref}[1]{Reference~\citen{#1}}

\newcommand{\figlabel}[1]{\label{fig:#1}}
\newcommand{\eqnlabel}[1]{\label{eqn:#1}}
\newcommand{\tablabel}[1]{\label{tab:#1}}

\renewcommand{\figref}[1]{Fig.~\ref{fig:#1}}
\renewcommand{\eqnref}[1]{Eq.~\ref{eqn:#1}}

\begin{document}                                                 

\begin{frontmatter}

\title{\boldmath Search for $\pkpidcsd$ and $\dskkpidcsd$ Using \GP Event Selection}

\collaboration{The~FOCUS~Collaboration}\footnotemark
\author[ucd]{J.~M.~Link}
\author[ucd]{P.~M.~Yager}
\author[cbpf]{J.~C.~Anjos}
\author[cbpf]{I.~Bediaga}
\author[cbpf]{C.~Castromonte}
\author[cbpf]{A.~A.~Machado}
\author[cbpf]{J.~Magnin}
\author[cbpf]{A.~Massafferri}
\author[cbpf]{J.~M.~de~Miranda}
\author[cbpf]{I.~M.~Pepe}
% ...

\address[ucd]{University of California, Davis, CA 95616}
\address[cbpf]{Centro Brasileiro de Pesquisas F\'\i sicas, Rio de Janeiro, RJ, Brazil}
% ...

\footnotetext{See \textrm{http://www-focus.fnal.gov/authors.html} for additional author information.}

\begin{abstract}  

We apply a genetic programming technique to search for the \dcs decays
\pkpidcsd and \dskkpidcsd. We normalize these decays to their \cf partners and
find $\text{BR}(\pkpidcsd)/\text{BR}(\pkpi) = (0.05 \pm 0.26 \pm 0.02)\%$ and
$\text{BR}(\dskkpidcsd)/\text{BR}(\dskkpi) = (0.52\pm 0.17\pm 0.11)\%$ where the
first errors are statistical and the second are systematic. Expressed as
$90\%$  confidence levels (CL), we find $< 0.46 \% $ and  $ < 0.78\%$
respectively. This is the first successful use of genetic programming in a high
energy physics data analysis.

\end{abstract}

\end{frontmatter}


Cabibbo suppressed (CS) and \dcs (DCS) decays are important in helping us
understand the dynamics of hadronic decay processes. DCS decays are unique to
the charmed hadrons; charm is the only heavy up-type quark that hadronizes. DCS
decay rates are such that only DCS decays of \dplus and \dzero have been
observed, while CS decays of nearly all the charmed hadrons have been observed.
This paper presents a search for DCS decays of \lc and \dsplus.   Both
branching ratios are expected to be small. Na\"{\i}ve expectations place DCS
branching ratios around $\tan^4 \theta_c$, or about 0.25\%, relative to their
\cf (CF) counterparts. Lipkin argues~\cite{Lipkin:2002za} that exact SU(3) symmetry would require the product
of the DCS relative branching ratios 
$\text{BR}(\dplus \to \kplus \piminus \piplus) /
\text{BR}(\dplus \to \kminus \piplus \piplus)$  and $\text{BR}(\dsplus \to
\kplus \kplus \piminus) / \text{BR}(\dsplus \to \kminus \kplus \piminus)$ to be
exactly $\tan^8 \theta_c$. This means the latter should  
be about 0.07\%; a much larger value requires a large
violation of flavor SU(3). In the \lc case, the CF normalizing mode has a \cq-\dq
\wplus exchange decay channel available, while the DCS decay mode may only
proceed through spectator decays. The lifetime difference between \lc and \xicp
shows us that this exchange mode is important, so we expect that the branching
ratio for \pkpidcsd should also be less than $\tan^4 \theta_c$. 

We have applied a genetic programming (GP)~\cite{Koza:gp1992} technique to
search for the DCS decays \dskkpidcsd and \pkpidcsd (charge-conjugate states are
implied), neither of which have been observed. GP is a machine learning
technique which evolves populations of programs (event filters in our case)
over a series of generations. The genetic programming learning mechanism is
modeled on biological and evolutionary principals and differs from some other
machine learning solutions in that the form of the solution is not
specified in advance but is determined by the complexity of the problem. A
full demonstration of this technique on the observed DCS decay \kpipidcsd is given
in \refref{Link:2005ns}. 

These results use data taken with the charm photoproduction experiment
FOCUS (FNAL-E831), an upgraded version of
FNAL-E687~\cite{Frabetti:1992au} which collected data using the Wideband photon
beamline during the 1996--1997 Fermilab fixed-target run.  The FOCUS experiment
utilizes a forward multiparticle spectrometer to study charmed particles
produced by the interaction of high energy photons  ($\langle E \rangle \approx
180~\gev$)~\cite{e687:nimbeam} with a segmented BeO target.  Charged particles
are tracked within the spectrometer by two silicon microvertex detector
systems. One system is interleaved with the target segments~\cite{Link:2002zg};
the other is downstream of the target region. These detectors provide excellent
separation of the production and decay vertices.  Further downstream, charged
particles are tracked and momentum analyzed with a system of five multiwire
proportional chambers~\cite{Link:2001dj} and two dipole magnets of opposite
polarity. Three multicell threshold \cer detectors are used to identify
electrons, pions, kaons, and protons. FOCUS also contains a
complement of hadronic and electromagnetic calorimeters and muon detectors.


We use loose analysis cuts on both DCS and CF decay modes to select
initial samples of events for
optimization by GP. The FOCUS \cer
algorithm~\cite{Link:2001pg} returns negative $2\times \text{log-likelihood}$
values $W_i(j)$ for particle $j$ and hypothesis $i \in {e, \pi, K, p}$.
Differences between log-likelihoods are used as particle ID, such as $\Delta
W_{Kp}(p) \equiv W_{K}(p) -  W_{p}(p)$ for ``proton favored over kaon.'' We require $\Delta W_{\pi K}(K) > 2$ for all kaons in both
decay modes. For protons from \lc candidates,
we require  $\Delta W_{\pi p}(p) > 4$ and  $\Delta W_{Kp}(p) > 0$ in the
initial selection. For the \lc, 
we also require that 
the separation between the production and decay vertices, $\ell$, is greater
than 3 times its error, $\sigma_\ell$.  For the
\dsplus, the vertex separation requirement is $\lsig > 6$. 
For both \lc and \dsplus, the three decay fragments must form a
vertex with a confidence level (CL)~$> 1\%$, and a production vertex is formed
by adding as many remaining tracks to the charm candidate as possible while
maintaining a vertex CL~$> 1\%$. One additional requirement is placed on the CF (DCS)
\dsplus candidates: the $K^-K^+\pi^+$ ($K^+K^+\pi^-$) combination is rejected if,
reconstructed as $K^-\pi^+\pi^+$ ($K^+\pi^+\pi^-$), the mass is within $2\sigma$
of the nominal \dplus mass. This cut removes a
prominent reflection from the CF candidates and stabilizes the many fits done
during the optimization process; it is applied to the DCS mode for consistency. The initial samples of \lc and \dsplus
candidates in CF and DCS decay modes are shown in \figref{lc_skim}.

\begin{figure}
\begin{center}
\includegraphics[width=7.0cm, bb=0 0 567 460]{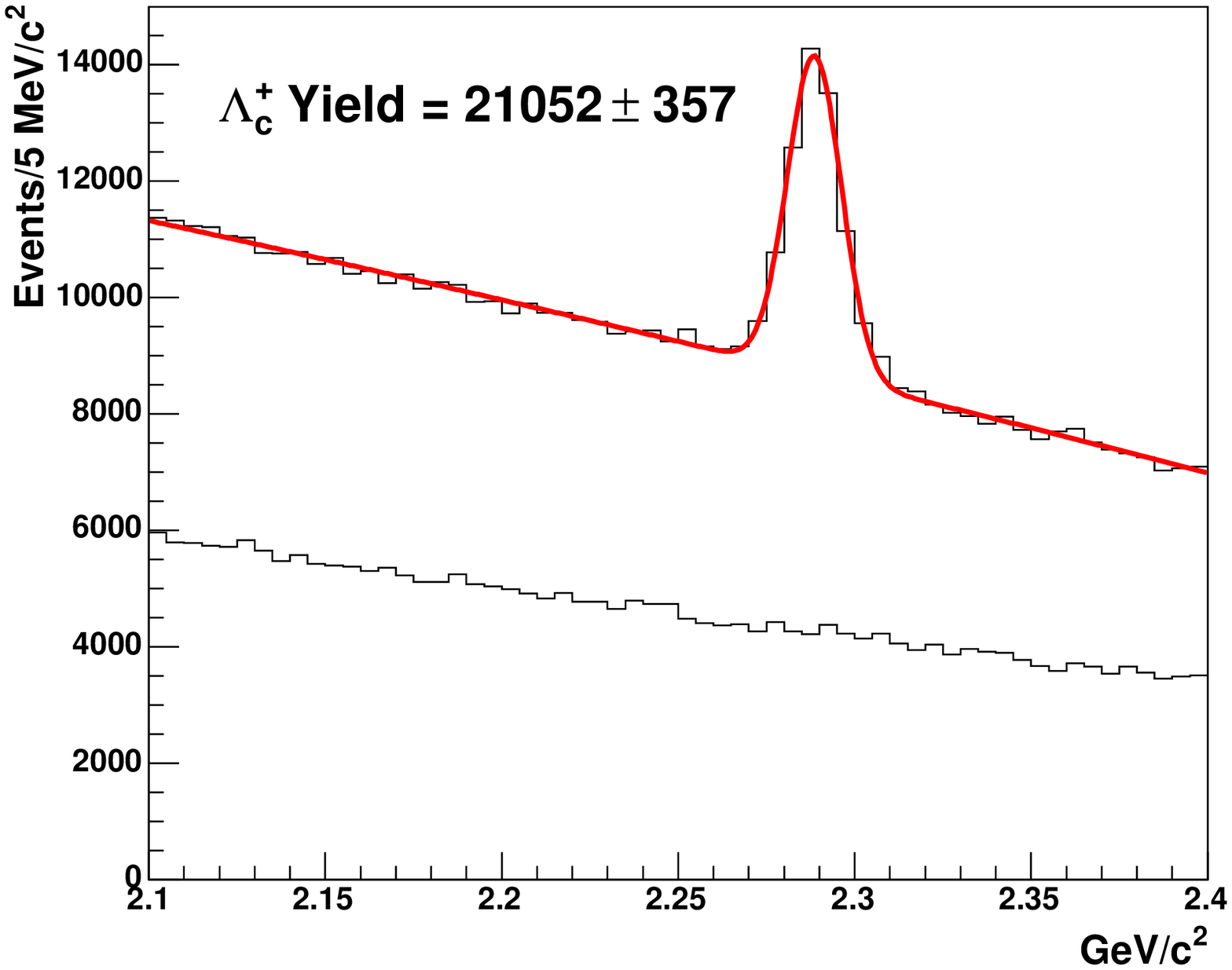}
\hspace*{-4mm}
\includegraphics[width=7.0cm, bb=0 0 567 460]{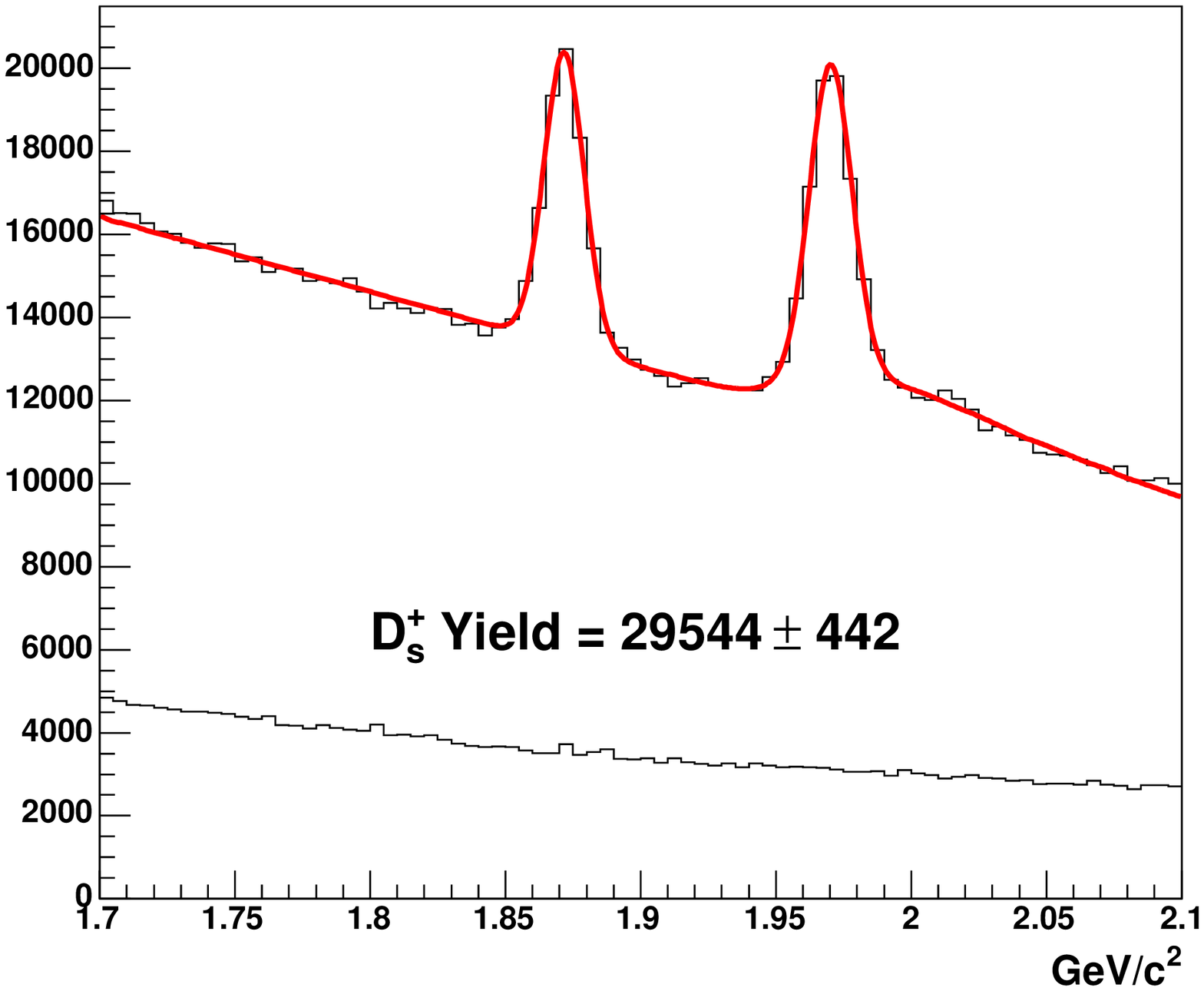}
\end{center}
\caption{Initial \lc (left) and \dsplus (right) data samples. The upper
distributions show the CF decay candidates, the lower distributions show the
DCS candidates. In the \dsplus plot, the \kpipi rejection cut described in the text
is \emph{not} applied and a contribution to the fit for this reflection is
visible to the right of the \dsplus peak.}
\figlabel{lc_skim}
\end{figure}

A GP framework (GPF) evolves and tests event filters. For each filter, we define
a fitness 
\begin{equation}
f \propto \frac{B_\text{DCS}}{S_\text{CF}^2} \times (1 +
0.005\times N_\text{nodes}) 
\eqnlabel{nosignal_fitness}
\end{equation}
where $B_\text{DCS}$ is the number of background events found in a fit of the
DCS mass distribution which excludes the signal region, and $S_\text{CF}$ is
the fitted CF yield. $S_\text{CF}/\sqrt{B_\text{DCS}}$ is proportional to the
projected DCS significance assuming no real DCS events and equal CF and DCS
selection efficiencies; squaring this quantity
further emphasizes ``better'' filters and inverting it allows small
fitnesses to describe good event filters. $N_\text{nodes}$ is the total number
of variables, constants, functions, and operators used in the filter and is included as a penalty
term to encourage smaller filters and to attempt to eliminate the addition of
nodes which do not select events based on physics. 

For the data samples in \figref{lc_skim}, half the events (as explained
later) along with a large number of variables (37 for \lc, 34 for \dsplus),
operators and functions (21), and constants are used as inputs to the GPF which
randomly generates a large number of filters and  calculates the fitness
of each. The GPF preferentially selects filters for which this
fitness is small to participate in breeding subsequent generations of filters.
In this way subsequent generations develop filters with better average
fitnesses. At the end of the process we use the filter with the single best
fitness to select events for further analysis.

The variables and resulting filter used in the CF and DCS decays are identical.
All variables commonly used in FOCUS analyses and some additional variables are
allowed to be used in the event filter. These can be roughly broken into
categories of vertexing, track quality, particle identification, production and
decay kinematics, away-side charm tagging and, for the \lc, evidence for decays
of the excited states $\Sigma_c^{(*)0,++}$. A description of the
variables\footnote{In addition to the variables described in
\refref{Link:2005ns}, we add three additional variables: $\Delta W_{K
\pi}(\pi)$, the number of tracks in the production vertex, and a value
indicating if any of the vertex detector track segments are shared between two
tracks.} used, examples of the event filters constructed, and how the
population of filters evolves over many generations can be found in
\refref{Link:2005ns}.  In both cases we use 20 sub-populations of 1500 event
filters per generation as described in \refref{Link:2005ns}.

When searching for \pkpidcsd and \dskkpidcsd decays (with the signal regions
masked), the GPF is allowed to run for 80 generations. The process is
terminated when no improvement in fitness is observed for about 10 generations.
The best \pkpidcsd filter found has 45 nodes and uses 12 unique physics
variables. The events selected are shown in \figref{lc_gp_sel}. One can see
that about 15\% of the signal is retained compared to \figref{lc_skim} while
the backgrounds are reduced by a factor of $\sim$1000. The distributions in
both the CF and DCS cases are fit with a second degree polynomial\footnote{No
significant reflections in \pkpidcsd or \dskkpidcsd are seen in high-statistics
MC studies (which include all known $\cq\cbq$ decay processes) of these decays, so we are justified in using simple background
shapes.} and a single Gaussian. In the DCS case, the Gaussian mean and $\sigma$
are fixed to the CF values and we find $1.2\pm6.6$ events. Correcting for the
relative efficiency $\epsilon_\text{CF}/\epsilon_\text{DCS} = 1.204 \pm 0.007$
(stat.) calculated with \mc (MC) simulations, we obtain a relative BR of
\begin{equation} 
\frac{\text{BR}(\pkpidcsd)}{\text{BR}(\pkpi)} = (0.05\pm 0.26)\% \, , 
\end{equation}  
which is consistent with zero.

\begin{figure}
\begin{center}
\includegraphics[width=13cm,bb=0 0 567 225]{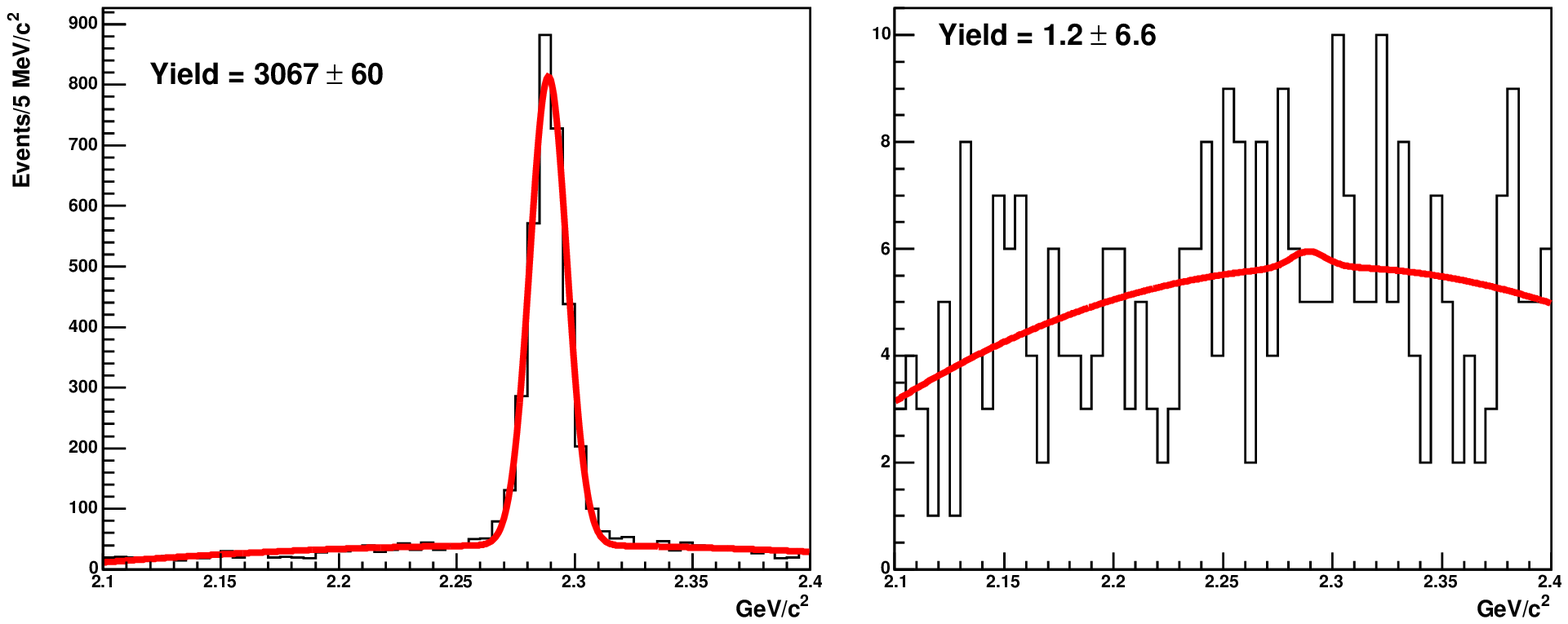}
\end{center}
\caption{\lc samples after selection. On the left is the CF normalizing mode, on
the right, the remaining \pkpidcsd candidates.}
\figlabel{lc_gp_sel}
\end{figure}

The best \dskkpidcsd filter found has 85 nodes and
uses 15 unique physics variables. The events selected are shown in
\figref{ds_gp_sel}. The fits shown are performed identically to the \lc case
except that an additional Gaussian is added to the
CF distribution for the CS decay \kkpi.  We find $27.5\pm9.2$ events in the
DCS distribution which, corrected by the relative efficiency
$\epsilon_\text{CF}/\epsilon_\text{DCS} = 1.154 \pm 0.005$, gives a
relative BR of
\begin{equation}
\frac{\text{BR}(\dskkpidcsd)}{\text{BR}(\dskkpi)} = (0.52\pm 0.17)\% \, .
\end{equation} 

\begin{figure}
\begin{center}
\includegraphics[width=13cm,bb=0 0 567 225]{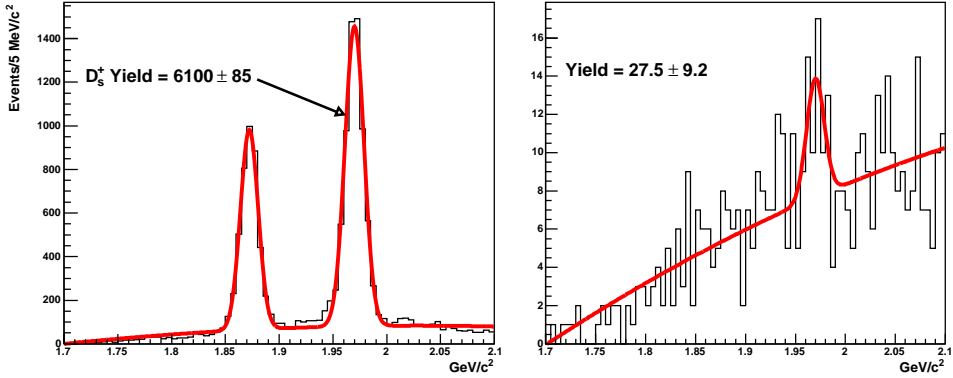}
\end{center}
\caption{\dsplus samples after selection. On the left is the CF normalizing
mode, on the right, the remaining \dskkpidcsd candidates.}
\figlabel{ds_gp_sel}
\end{figure}

In both cases our central values are calculated assuming non-resonant decays for
the DCS case and the best known resonance models for the CF decays (the
PDG~\cite{Eidelman:2004wy} model for \lc and a FOCUS model for \dsplus) as
explained below.


To convert these relative BRs into upper limits including systematic errors, we
use a method proposed by Convery~\cite{Convery:2003af} for incorporating
systemic uncertainties on reconstruction efficiencies into BR measurements when
a fit, rather than event counting, is used. In this case, the probability
$P(B)$ of the true BR being $B$ is given by 
\begin{equation}
 p(B) \propto
 \frac{1}{\sqrt{\frac{B^2}{2\sigma_B^2}+\frac{\hat{S}^2}{2\sigma_S^2}}} 
\, \exp \left[\frac{-(B-\hat{B})^2}{2\left(\frac{B^2\sigma_S^2}{\hat{S}^2} +
\sigma_B^2 \right)}\right]
\eqnlabel{br_prob}
\end{equation}
where $\hat{B}$ is the fitted BR, $\sigma_B$ is its error and $\sigma_S/\hat{S}$ is
the percent systematic error on the efficiency. $P(B)$ is numerically
integrated until the point at which 90\% of the physical $(\textrm{BR} >0)$ area is
included. This point is reported as the 90\% confidence level. If 
$\sigma_S/\hat{S} \gg 10\%$, this distribution has a long high-end tail,
raising the 90\% limit considerably.  

We consider four sources of systematic error on our knowledge of the relative
efficiencies of the CF and DCS decay modes. First, and negligible, is the
number of MC events used. Second and third, we consider the effects of
different resonance models for the DCS and CF states respectively. Finally, we
consider whether the evolved event selector may have different efficiencies for
the CF and DCS modes.

In studying possible resonances for \pkpidcsd candidates, we calculate 
efficiencies as if the final state is entirely non-resonant or entirely $\lc
\to \Delta(1232)^{0} \kplus$ or $\lc \to \proton \kstar$. The systematic error
is taken as the standard deviation of the three possible efficiencies. For
the \dskkpidcsd candidates, we consider non-resonant decays,  $\dsplus \to
\kstar \kplus$, and $\dsplus \to \kfttr \kplus$ in the same way. From these
studies we find 5.3\% and 10.7\% systematic uncertainties on the \lc and \dsplus 
efficiencies respectively.

The resonant structures of the CF decays are reasonably well known. For the
\lc, we use two models, one from the PDG and another which excludes the
$\Lambda(1520)^{0} \piplus$ decay mode. For the \dsplus, we consider an
incoherent model based on the PDG averages and a coherent
model~\cite{Malvezzi:2002xt} developed from the FOCUS data. From these studies
we find 2.1\% and 2.6\% uncertainties on the \pkpi and \dskkpi efficiencies,
respectively.

Our final systematic contribution is motivated by the concern that the
efficiency of the final GP-generated event filter may differ for the CF and DCS
modes (after correction for kinematic acceptance of different final states) in a
way that is not well modeled by MC. Since this is impossible to measure, we
adopt a more rigorous test. We test if the event filter has the same
efficiency on CF data and MC events. We do this by comparing the CF yields of
data and MC samples before and after the event filter is applied.\footnote{The
\kpipi rejection cut is applied with the event filter in the \dsplus case.} For the \lc,
we find that the event filter retains $14.5\pm0.4\%$ and $14.9\pm 0.1\%$ of the data
and MC events, respectively. For the \dsplus, we determine these quantities to
be $21.0\pm0.4\%$ and $20.3\pm 0.1\%$. We take the differences between these numbers
(neglecting the errors) as systematic uncertainties; these cause 2.6\% and 3.5\%
uncertainties on the relative efficiencies for \pkpidcsd and
\dskkpidcsd respectively. All systematic uncertainties on the relative
efficiencies are summarized in \tabref{syst_summ}.

\begin{table}
\caption{Summary of systematic uncertainties. Listed are the percent
uncertainties on the relative efficiencies of the DCS and CF decay modes from
various sources.}
\tablabel{syst_summ}
\begin{center}
\begin{tabular}{lrr}\hline
 & \multicolumn{2}{c}{Syst.\ Unc.\ (\%)} \\
\multicolumn{1}{c}{Source}          & \lc & \dsplus \\ \hline
MC statistics   & 0.6 & 0.4 \\
DCS resonances  & 5.3 & 10.7 \\
CF resonances   & 2.1 & 2.6 \\
GP filter       & 2.6 & 3.5 \\\hline
Total           & 6.3 & 11.6 \\\hline
\end{tabular}
\end{center}
\end{table}

Finally, as mentioned above, we only use half (even-numbered) of the events in
the optimization of the event filter. The final values use the event filter
applied to all the events, but as a check, we divide the sample into events the
GPF used and did not use. We measure the BR independently for these two samples
and see no significant evidence for a difference, strongly suggesting that the
GPF is not arbitrarily selecting or rejecting small numbers of events to
artificially reduce backgrounds or enhance signals.


Using the total percent errors in \tabref{syst_summ} as $\sigma_S/\hat{S}$ and
the above BRs, statistical errors, and percent systematic errors, we integrate
$P(B)$ from \eqnref{br_prob} as described
and find  
\begin{equation}
\frac{\text{BR}(\pkpidcsd)}{\text{BR}(\pkpi)} < 0.46\%
\end{equation} 
and
\begin{equation}
\frac{\text{BR}(\dskkpidcsd)}{\text{BR}(\dskkpi)} < 0.78\%
\end{equation} 
where the limits are at the 90\% CL.
We also determine effective systematic uncertainties for our measurements by
calculating the uncertainty necessary, when added in quadrature to the
statistical uncertainty, to cover the central 68\% of the distrubution in
\eqnref{br_prob}. By this method, we find
$\text{BR}(\pkpidcsd)/\text{BR}(\pkpi) = (0.05 \pm 0.26 \pm 0.02)\%$ and
$\text{BR}(\dskkpidcsd)/\text{BR}(\dskkpi) = (0.52\pm0.17\pm0.11) \% $ where the
first errors are statistical and the second are systematic. 
The distributions described by
\eqnref{br_prob} and the 90\% integrals for the DCS decays are shown in
\figref{br_integral}. Both limits are larger than the expected
($\lesssim \tan^4 \theta_c$) level, but are the first reported limits on
these decays. Furthermore, this is the first successful application of the GP
technique to an HEP data analysis.

\begin{figure}
\begin{center}
\includegraphics[width=6.5cm,bb=30 30 567 450]{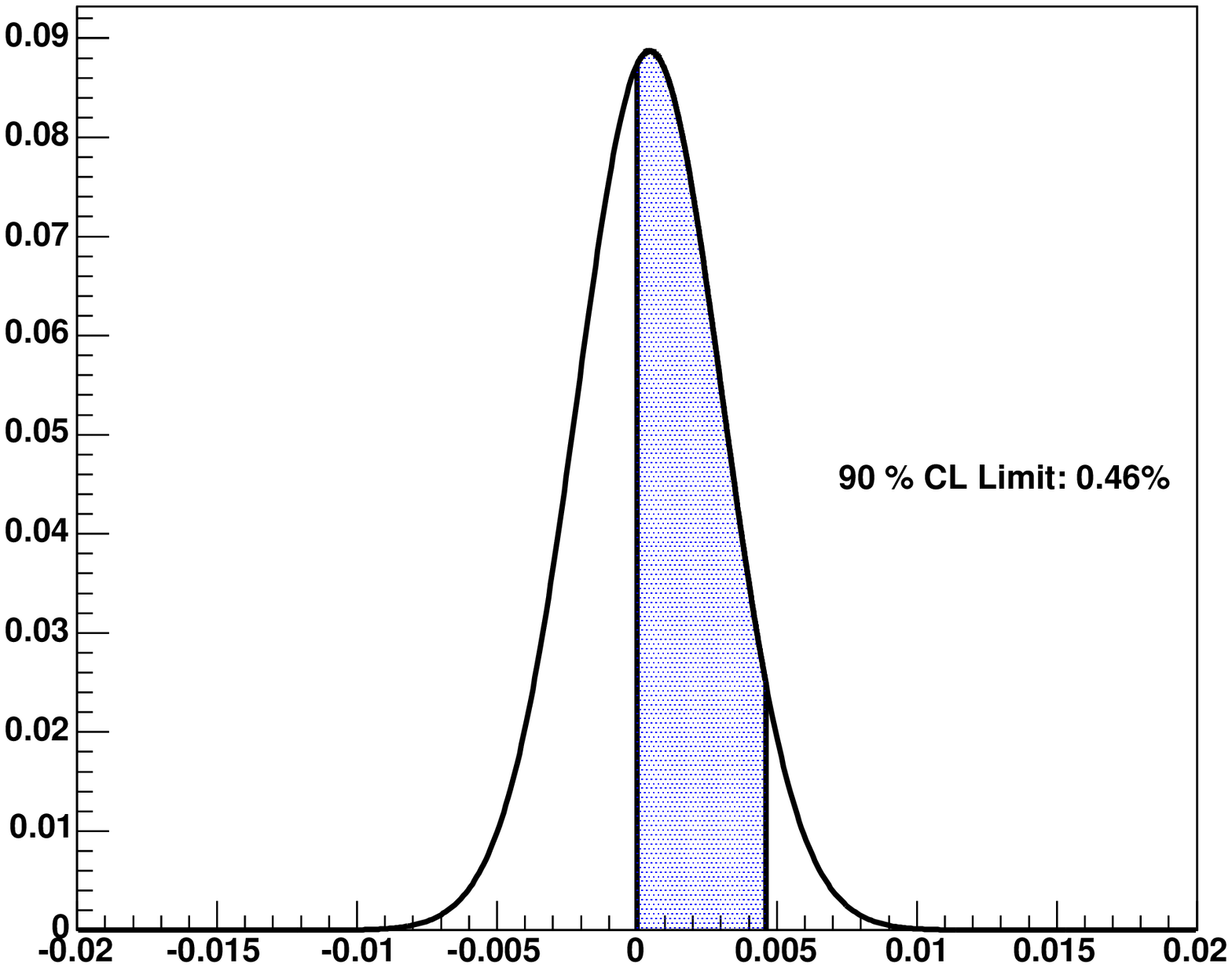}
\hspace{3mm}
\includegraphics[width=6.5cm,bb=30 30 567 450]{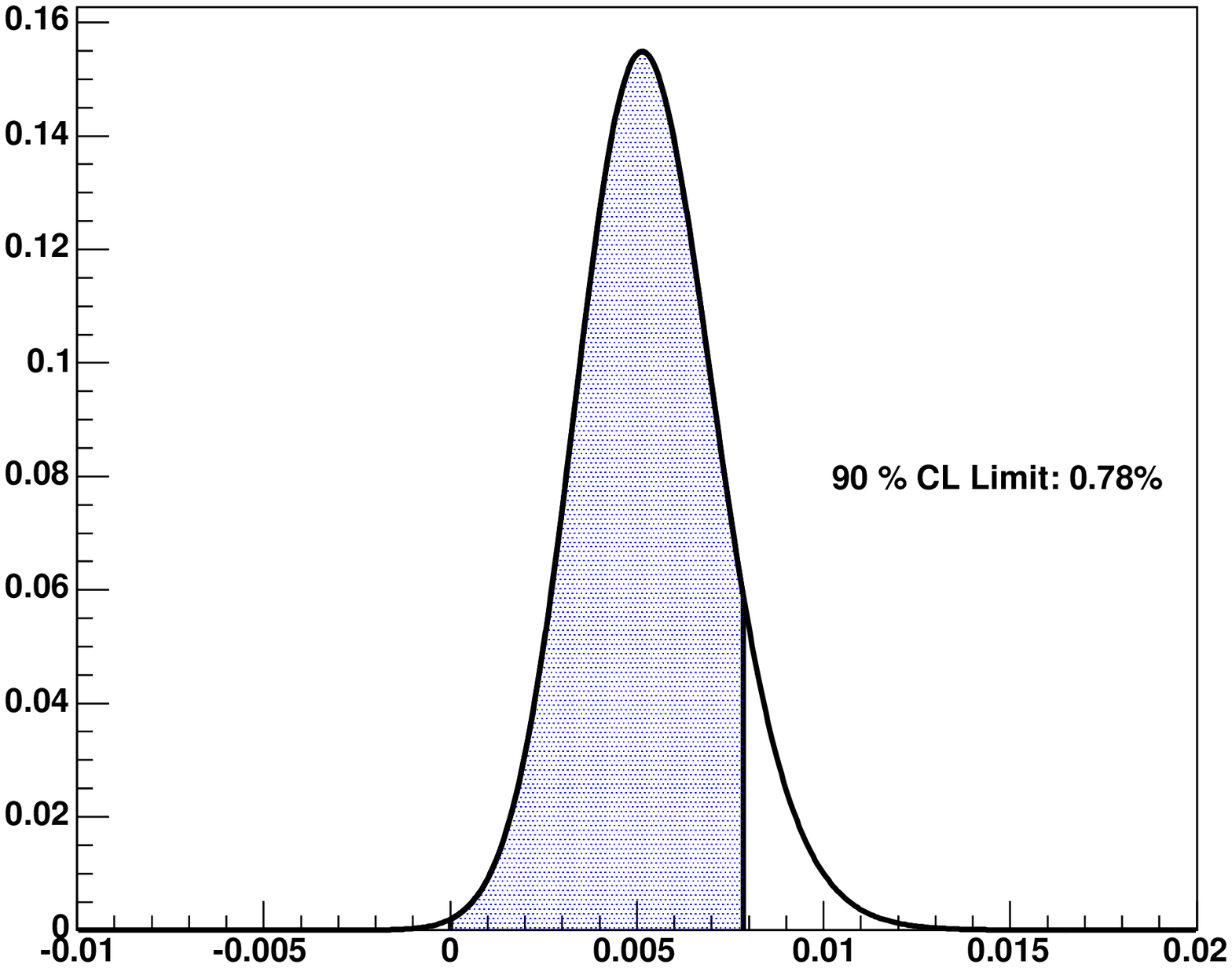}
\end{center}
\caption{Relative BR limit determination for \pkpidcsd (left) and \dskkpidcsd (right). 
The curves show the branching ratio probability for DCS decays
relative to CF decays. The vertical axes are arbitrary. The shaded areas show the
90\% integrals over the physical range.}
\figlabel{br_integral}
\end{figure}

\section*{Acknowledgments}

We wish to acknowledge the assistance of the staffs of Fermi National
Accelerator Laboratory, the INFN of Italy, and the physics departments
of the collaborating institutions. This research was supported in part
by the U.~S.  National Science Foundation, the U.~S. Department of
Energy, the Italian Istituto Nazionale di Fisica Nucleare and
Ministero dell'Istruzione dell'Universit\`a e della Ricerca, the
Brazilian Conselho Nacional de Desenvolvimento Cient\'{\i}fico e
Tecnol\'ogico, CONACyT-M\'exico, the Korean Ministry of Education, 
and the Korean Science and Engineering Foundation.


\bibliographystyle{elsart-num}
\bibliography{physjabb,abbrev,pdg,focus,e687,theory,gp,other}

\end{document}